\documentstyle[epsfig,12pt,a4p,subfigure]{article}

\newcommand{\etpim}{\mbox{$\eta\pi^-$ }}
\newcommand{\be}{\begin{equation}}
\newcommand{\ee}{\end{equation}}
\newcommand{\bea}{\begin{eqnarray}}
\newcommand{\eea}{\end{eqnarray}}
\newcommand{\bean}{\begin{eqnarray*}}
\newcommand{\eean}{\end{eqnarray*}}

\newcommand{\pom}{$I\hspace{-1.6mm}P$}

\newcommand{\gapproxeq}{\lower
.7ex\hbox{$\;\stackrel{\textstyle >}{\sim}\;$}}
\newcommand{\lapproxeq}{\lower
.7ex\hbox{$\;\stackrel{\textstyle <}{\sim}\;$}}

\begin{document}
\begin{titlepage}
\begin{tabbing}
wwwwwwwwwwwwwwwright hand corner using tabbing so it looks neat and in \= \kill
\> {8 August 2000}
\end{tabbing}
\baselineskip=18pt
\vskip 0.7in
\begin{center}
{\bf \LARGE Isospin breaking
exposed in
$f_0(980)- a_0(980)$ mixing}\\
\vspace*{0.9in}
{\large Frank E. Close}\footnote{\tt{e-mail: F.E.Close@rl.ac.uk}} \\
\vspace{.1in}
{\it Rutherford Appleton Laboratory}\\
{\it Chilton, Didcot, OX11 0QX, England}\\
\vspace{0.1in}
{\large Andrew Kirk}\footnote{\tt{e-mail: ak@hep.ph.bham.ac.uk}} \\
{\it School of Physics and Astronomy}\\
{\it Birmingham University}\\
\end{center}
\begin{abstract}
We suggest that mixing between the $f_0(980)$ and $a_0(980)$,
due to their dynamical
interaction with the nearby $K\bar{K}$ thresholds, can give rise
to a significantly enhanced production rate of $a_0(980)$ relative
to $a_2(1320)$ in
$pp \rightarrow p_{s} (\eta \pi^o) p_{f}$ as $x_F \rightarrow 0$.
The peaking
of the cross section as $\phi \to 0$ should also occur.
We show that such effects are
seen in data and deduce that the $f_0(980) - a_0(980)$ mixing
intensity is 8~$\pm$~3~\%.
\end{abstract}
\end{titlepage}
\setcounter{page}{2}
\par
The enigma of the scalar mesons may be boiled down to an
essential question: what are the $f_0(980)$ and $a_0(980)$?
Do they have a common origin and, if so, what is it?
Understanding the $f_0(980)$ in particular is a central problem
for identifying the dynamics associated with the long sought
scalar glueball.
\par
There have even been suggestions that the $f_0(980)$ itself
may be the eponymous glueball, perhaps mixed with $q \bar{q}$;
in such a case the mass degeneracy with the $a_0(980)$ would be
somewhat accidental and the two mesons not clearly related.
An interpretation of the $f_0(980)$
as a $q \bar{q}$ state
is still
consistent with the present data (see for example ref.~\cite{montanet}).
By contrast, there is a large body of work drawing on the
observation that the $f_0(980)$ and $a_0(980)$
are very close to the $K \bar{K}$ threshold, and that
the $K \bar{K}$ channel drives the dynamics \cite{tornqv}.
As an extreme, there is the possibility that these mesons
are truly bound states of $K \bar{K}$ \cite{iswein}.
\par

Traditionally in strong interactions isospin is
believed to be a nearly exact symmetry, broken only by
the slightly different masses of the $u$ and $d$ quarks and/or
electroweak effects. The small difference in mass between $K^{\pm}$
and $K^0$ is a particular example. However, the mass gaps between
the $f_0(980)/a_0(980)$ and the $K^+K^-$ and $K^0K^0$ thresholds
are substantially different with the result that the dynamics of
bound $K \bar{K}$ states can be described better in a basis
specified by mass eigenstates. Such dynamics would give rise to
a violation of isospin and lead to mixing of states with
different G-parities.

\par
The possibility of such an effect was suggested long ago
 in ref.~\cite{achasov1}.
In ref.~\cite{achasov2} a study was performed of the production
of the $a_0(980)$ in the reaction $\pi^+\pi^- \rightarrow \eta \pi$
which due to G parity is forbidden and can only occur through
$f_0(980)-a_0(980)$ mixing. This showed that (6--33)\%
of the $a_0(980)$ cross section in $\pi^-p$ reactions
could be due to
$f_0(980)-a_0(980)$ mixing. Further discussions along this line have
been made by ref.~\cite{speth} who have specifically drawn attention
to the relation between the existence of $K\bar{K}$ molecular bound
states and large violations of isospin. Very recently, attention
has been drawn to such mixings having observable effects in
threshold photoproduction, such as at CEBAF~\cite{kerbikov}.
These papers have all concentrated on the
production of the $f_0(980)/a_0(980)$ by flavoured mesons or photons;
in this paper we propose that their production by gluonic systems,
such as the \pom (Pomeron)-induced
production in the central region at high energy:
$pp \to pp + f_0(980)/a_0(980)$, may provide rather clean tests of the mixing.
Furthermore, we shall suggest that new data from the WA102 collaboration
at CERN~\cite{etapipap}
are already consistent with a significant mixing. We shall
consider alternative interpretations and suggest ways of eliminating these
in future experiments.
\par
These data potentially may help to elucidate the
nature of the $f_0(980) / a_0(980)$ states. Our hypothesis is
based on recent breakthroughs in understanding the dynamics
and topology (momentum and spatial distributions) of meson
production in the central region of rapidity, $pp \to pMp$~\cite{cs,cks}.
In particular, we shall focus on
the description of the observed $\phi$ dependences~\cite{cks},
where $\phi$ is the angle between the $p_T$ vectors of the two outgoing
protons.
In such processes at high energy, where \pom \pom \thinspace \thinspace fusion
dominates the meson production, $C=+,I=0$ resonances such as
the $f_0(980)$ are very strongly produced~\cite{pipikkpap}
whereas in general isospin 1 states are suppressed~\cite{pipipap}.
Even at the energies of the WA102 data, there is considerable evidence
that \pom \pom \thinspace \thinspace fusion
is an important part of the production
dynamics~\cite{pipipap}.
It is tantalising therefore that recent data from the
 WA102 collaboration on the
 centrally produced $\eta \pi $ final state~\cite{etapipap} show
interesting effects in that they
are in accord with substantial $f_0(980) - a_0(980)$
mixing.
\par
In particular it is instructive to compare the systematics
of the well understood $f_2(1270)/a_2(1320)$ $(^3P_2 q\bar{q})$
states with the
$f_0(980)/a_0(980)$ states.
In the reaction $pp \rightarrow p (\eta \pi^o) p$
the centrally produced $a_0(980)$ and $a_2(1320)$ are suppressed
relative to their $I=0$ partners, as expected
for  $I=1$
states. Nonetheless, there appears to be an extra affinity for
 $a_0(980)$ production here, since

\begin{equation}
\frac{\sigma (pp \rightarrow pp[a_0^0(980) \rightarrow \eta \pi])}
{\sigma (pp \rightarrow pp[a_2^0(1320) \rightarrow \eta \pi])}
\approx 2.0 \pm 0.4
\label{eq:r1}
\end{equation}


\noindent By contrast, when the charged members of these isovectors
are produced,
 as in $pp \rightarrow p (\eta \pi^-) \Delta^{++} $,
$a_0^-(980)$ and $a_2^-(1320)$ production rates are found to be similar.
Fits to the \etpim mass spectrum
in central production give
\begin{equation}
\frac{\sigma (pp \rightarrow p\Delta^{++}[a_0^-(980) \rightarrow \eta \pi^-])}
{\sigma (pp \rightarrow p\Delta^{++}[a_2^-(1320) \rightarrow \eta \pi^-])}
\approx 0.8 \pm 0.2
\label{eq:r2}
\end{equation}
\noindent The significance of these ratios becomes more apparent when
compared with
the case of the charge exchange reaction, where (as in eq.~(2))
$I=1$ exchanges are necessarily
present. In this case the $a_2(1320)$
meson dominates
the mass spectrum, and the ratio
\begin{equation}
\frac{\sigma (\pi^-p \rightarrow [a_0(980) \rightarrow \eta \pi]n)}
{\sigma (\pi^-p \rightarrow [a_2(1320) \rightarrow \eta \pi]n)}
\approx 0.15
\label{eq:r3}
\end{equation}
\noindent at 38 GeV/c beam momentum.
\par
First we shall explain this hierarchy and motivate the enhancement
in  (\ref{eq:r1}) as indicative of direct $f_0(980)$ production with
 $f_0(980) - a_0(980)$ mixing. Then we show how the characteristic momentum and
$\phi$ dependences of $f_0(980)$ production will, through mixing,
spill over to $a_0(980)$ production. Finally we shall see that
such signatures are indeed
present in the $a_0(980)$ production data and consistent with a
substantial $f_0(980)-a_0(980)$ mixing.
\par
In $\pi^- p \to a_{0,2}n$ (\ref{eq:r3}), it is  easy to make the
$a_2(1320)$ via $\rho$ exchange. However in order to produce the
$a_0(980)$, $\rho_2$ and/or $b_1$ exchange is needed which is relatively
suppressed~\cite{achasov2}. In $pp \rightarrow p (\eta \pi^-) \Delta^{++}$
(\ref{eq:r2})
the $a_2$ production is again consistent
with $\pi \rho$ fusion~\cite{etapipap}.
Fig.~\ref{fi:1}b) shows the observed $\phi$ distribution for the
$a_2^-(1320)$~\cite{etapipap}. As can be seen the
distribution is
isotropic in $\phi$,
as expected for $\pi$ exchange~\cite{cs,Diehl},
and the $t$ slopes~\cite{etapipap} are
consistent with $\pi$ and $\rho$ being produced at either vertex, (it is
known that the $\rho$ can be produced
at the $p \Delta^{++}$ vertex from the WA102
data on $pp \to p \Delta^{++} \rho^-$ \cite{phiangpap}). However, some
other mechanism is needed to explain the relatively enhanced $a_0(980)$ signal.
The $\phi$ distribution for the $a_0^-(980)$ is shown in fig.~\ref{fi:1}a) and
as can be seen it is also isotropic.
\par
There are four particular exchanges that can enhance the $a_0(980)$ signal
in $pp \rightarrow p (\eta \pi^-) \Delta^{++}$ (\ref{eq:r2})relative to its
suppressed rate in charge exchange (\ref{eq:r3}).
First, I=0 exchange ($\eta$) can
occur at the proton vertex and cause $\pi \eta \to a_0(980)/a_2(1320)$.
Though $\eta$ exchange will be isotropic in $\phi$, in accord with data,
it is generally agreed to be small  and hence unlikely on its own
to drive the
enhanced $a_0(980)$ signal.
\par
The second possibility is production by $\pi b_1$ fusion.
 Although the $ppb_1$ vertex is
small, for $p\Delta b_1$ the quantum numbers match in $S-$wave and so
$\pi b_1$ fusion could be significant in $pp \rightarrow p \Delta a_{0,2}$.
Because of the $\pi$ exchange, the $\phi$ distribution will be
isotropic~\cite{cs,Diehl}, as in the data. However, empirically
$\sigma(pp \rightarrow ppa_2(1320))
\sim \sigma(pp \rightarrow p \Delta a_2(1320))$
which suggests that $b_1$ exchange is not the major mode
and further points to
$\pi \rho \to a_2(1320)$ as the dominant dynamics.
If $a_0(980) =$ $^3P_0(q\bar{q})$ then
in the quark model the ratio of amplitudes
$\pi b_1 \to a_0(980)/a_2(1320) \sim 1$
and we would still be left with the mystery
of its
production. Even if $a_0(980) \neq$ $^3P_0(q\bar{q})$, the $\pi b_1$ production
would be expected to be minimal in $pp$ and so the enigma of $a_0(980)$
production there would remain.
\par
 The third possibility is that $\rho$ from the $p \Delta$
vertex fuses with $\omega$ from the $pp$ vertex. This can feed both $a_0(980)$
and $a_2(1320)$.
Empirically the $a_2(1320)$ is produced polarised with $\lambda = 1$
\cite{etapipap}; however, $VV \rightarrow 2^{++}(\lambda = 1)$ would
contain a characteristic $\sin^2 (\phi/2)$ component~\cite{cs}
in marked contrast
to the observed isotropy. This suggests that $\rho \omega \to a_2(1320)$ is not
a major mechanism and to the extent that $a_{0,2}$ are related as
$^3P_{0,2}$ $q\bar{q}$ states, would also argue against a
strong $a_0(980)$ signal.
Furthermore, the empirical absence of $a_2(1320) (^3P_2q\bar{q})$
with $(\lambda = 0)$ would in turn also imply a suppressed
production of $a_0(980) (^3P_0q\bar{q})$. However, it is possible that the
$K\bar{K}$ threshold disturbs the $a_0(980)$ such that $\rho \omega \to
a_0(980)$ is controlled by this and not by the $q \bar{q}$ content; in this
case
the production strength and properties could be independent of the $a_2(1320)$.
In general the $\phi$ dependence
for a $0^{++}$ state produced by vector-vector fusion
(where $L$ is the the longitudinal component of the vector and
$T$ is the transverse component)
has the following structure~\cite{cks}:

\begin{equation}
\frac{d\sigma}{dt_1 dt_2 d \phi}
\sim
[1 +
\frac{\sqrt{t_1t_2}}{\mu^2}\frac{a_T}{a_L}e^{(b_L-b_T)(t_1+t_2)/2 }
\cos(\phi)]^2 e^{-b_L(t_1+t_2) }
\label{eq:b}
\end{equation}

\noindent The ratio
$a_T/a_L$, which determines the relative importance
of the $0^+$ production by $T$ or $L$ components,can be positive or
negative, or in general even complex; its value is
determined, inter alia, by the internal dynamics of
the produced meson.
To the extent that the $\phi$ distributions
empirically are consistent with being isotropic, it would appear that
longitudinal-scalar amplitudes
dominate the production for the $a_0(980)$;
this might be natural were it a $K\bar{K}$
molecule where $K$ exchange dominated the production vertex.
\par
The fourth possibility is that \pom \thinspace \thinspace exchange
plays a role at the $pp$ vertex.
In principle there
could be significant $a_{0,2}$ \pom $\rightarrow a_{0,2}$.
If these were dominant,
one would expect similar production
rates of $a_0(980)$ in both $ppa_0(980)$ and $p\Delta a_0(980)$ processes
and also a rapid fall off in the $a_0(980)/a_2(1320)$ production ratio
with increasing energy.
As the data are only at a single value of $s$ one cannot immediately eliminate
this. However there are two features that argue against this.
First, $a_0(980)$ \pom $\rightarrow a_0(980)$ will give an isotropic $\phi$
distribution;
while this is seen in the $p \Delta a_0(980)$ production (fig.~\ref{fi:1}a),
the reaction $pp a_0(980)$ is $\phi$ dependent (fig.~\ref{fi:2}a).
Second; the $x_F$ distributions of the
$a_0^-(980)$ and $a_2^-(1320)$ formed in $p \Delta a_{0,2}$ are shown in
fig~\ref{fi:1}c) and d) respectively.
As can be seen the distributions are flat for
$x_F \leq 0.1$ (do not peak as $x_F \rightarrow 0$) which
may indicate that there is
a significant presence of non-central production.
Fig~\ref{fi:2}c) and d) show the $x_F$ distributions for the
$a_0^0(980)$ and $a_2^0(1320)$ formed in $p p a_{0,2}$. The distribution for
the $a_2^0(1320)$ is similar to that observed for the $a_2^-(1320)$
whereas that for the $a_0^0(980)$ is significantly different and peaks at
$x_F = 0$.
Indeed this is the only state with $I=1$ that is observed
to have a $x_F$ distribution peaked at zero~\cite{sumpap},
and moreover the distribution for the $a_0^0(980)$
looks similar to  the central production of states that are accessible to
\pom \pom \thinspace fusion,
in particular \pom \pom $\rightarrow f_0(980)$, see
fig~\ref{fi:3}c) and d). If we restrict ourselves to the central production
region $x_F \leq 0.1$, then the relative ratio of $a_0/a_2$ production
rates  in eq. (1) is even more enhanced and becomes
$3.4 \pm 0.4$ .
\par
In summary, we are unable to find an explanation of the production
of $a_0$ in $pp \rightarrow p\Delta a_0$ if $a_0 =$ $^3P_0q\bar{q}$. We will
now show evidence that there is significant mixing between
$a^0_0(980)$ and $f_0(980)$ in $pp \rightarrow pp a_0/f_0$, which
reveals a marked affinity of these states for $K\bar{K}$.
\par
In the process  $pp \rightarrow p (\eta \pi^0) p$ (\ref{eq:r1}),
there is a prominent new feature allowed, namely
\pom \pom \thinspace fusion
due to \pom \thinspace emission at each proton vertex. As
this will feed only $I=0$ channels, such as the $f_0(980)$ and $f_2(1270)$,
one would not expect this to affect $a_{0,2}$ production
unless isospin is broken.
As we noted earlier, the $a_0(980)/a_2(1320)$ ratio in the WA102 data
is significantly larger in reaction~(\ref{eq:r1})
than in reaction~(\ref{eq:r2}),
especially so when $x_F \leq 0.1$.
Furthermore, the $x_F$ distribution of the $a_0(980)$
production is, within the errors, identical to that of the
$f_0(980)$ (see fig.~\ref{fi:3}d)).
In reaction~(\ref{eq:r2}) the $\phi$ dependencies for both the
$a_0(980)$ and $a_2(1320)$ are flat (fig.~\ref{fi:1}a) and b) respectively).
In reaction~(\ref{eq:r1})
although the $\phi$ dependence of the $a_2(1320)$ remains
flat~(fig.~\ref{fi:2}b)) that
of the $a_0(980)$ is peaked as $\phi \rightarrow 0$~(fig.~\ref{fi:2}a)).
In fact the $\phi$ distribution for the $a_0(980)$ looks
very similar to that observed for the $f_0(980)$~(fig.~\ref{fi:3}a) and c)).
Qualitatively this is what would
be expected if
part of the centrally produced $a_0^0(980)$ is due to
\pom \pom $\to f_0(980)$ followed by
mixing between the $f_0(980)$ and the $a_0(980)$.
\par
In order to estimate the amount of the $a_0^0(980)$
that has been produced by mixing we
have performed a fit to the $\phi$ distribution of the $a_0^0(980)$
assuming it to be the sum of two incoherent
components: (i) a flat distribution similar to the
$a_0^-(980)$ and (ii) a distribution of the form $(4+\cos(\phi))^2$
which describes the $\phi$ distribution of the
$f_0(980)$ as shown in fig~\ref{fi:3}a).
We have determined from the fit to
fig.~\ref{fi:2}a) that 80~$\pm$~25~\% of the $a_0^0(980)$ comes from the
$f_0(980)$.
Combining this result with the relative total cross sections for the production
the $f_0(980)$ and $a_0^0(980)$~\cite{sumpap} we find the
$f_0(980)-a_0(980)$ mixing intensity to be 8~$\pm$~3~\%.
\par

\par
Technically our analysis only sets an upper limit on the isospin breaking
until such time as the energy dependence
is determined and the \pom \pom \thinspace
production thereby confirmed. Subject to this
caveat our analysis adds weight to the hypothesis that the
$f_0(980)$ and $a_0(980)$
are siblings that strongly mix, and that the $a_0(980)$ is not simply
a $^3P_0 q\bar{q}$ partner of
the $a_2(1320)$.
A natural explanation of these results would be that
the $K\bar{K}$ threshold plays an essential role in the existence and
properties of these states.
The question of whether they are $K \bar{K}$ bound states or whether
it is merely the $K \bar{K}$ threshold which is driving these effects
is still to be resolved.
\par
Other lines of study are now warranted.
Experimentally to confirm these ideas requires measuring the
production of the $\eta\pi$ channel at a much higher energy, for
example, at LHC, Fermilab
or RHIC where Reggeon exchanges such as $\rho \omega$
would be effectively zero
and hence any $a_0(980)$ production must come from isospin breaking
effects.
In addition, ``pure" flavour channels should now be
explored. Examples are $D_s$ decays~\cite{lipkin}
where the weak decay leads to a pure I=1 light hadron final state. Thus
$\pi f_0(980)$ will be
(and is~\cite{DS}) prominent, while our analysis would suggest
that $\pi a_0$ should also be present at $8 \pm 3$ \% intensity. We
recommend that these be studied with high statistics data sets now emerging
from E791, FOCUS and BaBar.
In addition, we encourage studies of $J/\psi$ decays at Beijing,
 in particular to the
``forbidden" final states $\omega a_0$ and $\phi a_0$ where we predict
branching ratios of $O(10^{-5})$.
On the theory side, detailed predictions are needed in specific
models in order to resolve precisely how the $K\bar{K}$ threshold
relates to the $f_0(980)/a_0(980)$ states.
\begin{center}
{\bf Acknowledgements}
\end{center}
\par
We are indebted to H.J.Lipkin for comments on heavy flavour decays.
This work is supported, in part, by grants from
the British Particle Physics and Astronomy Research Council,
the British Royal Society,
and the European Community Human Mobility Program Eurodafne,
contract NCT98-0169.

\clearpage
{ \large \bf Figures \rm}
\begin{figure}[h]
\caption{
For the reaction $pp \rightarrow \Delta^{++}p \eta \pi^-$:
The $\phi$ distributions for
a) the $a_0^-(980)$ and
b) the $a_2^-(1320)$.
The $x_F$ distributions for
c) the $a_0^-(980)$ and
d) the $a_2^-(1320)$.
}
\label{fi:1}
\end{figure}
\begin{figure}[h]
\caption{
For the reaction $pp \rightarrow pp \eta \pi^0$:
The $\phi$ distributions for
a) the $a_0^0(980)$ and
b) the $a_2^0(1320)$.
The $x_F$ distributions for
c) the $a_0^0(980)$ and
d) the $a_2^0(1320)$.
}
\label{fi:2}
\end{figure}
\begin{figure}[h]
\caption{
The $\phi$ distributions
a) for the reaction $pp \rightarrow pp f_0(980)$ and
b) for the $f_0(980)$ compared to the $a_0^0(980)$.
The
$x_F$ distributions
c) for the reaction $pp \rightarrow pp f_0(980)$ and
d) for the $f_0(980)$ compared to the $a_0^0(980)$.
}
\label{fi:3}
\end{figure}
\newpage
\begin{center}
\epsfig{figure=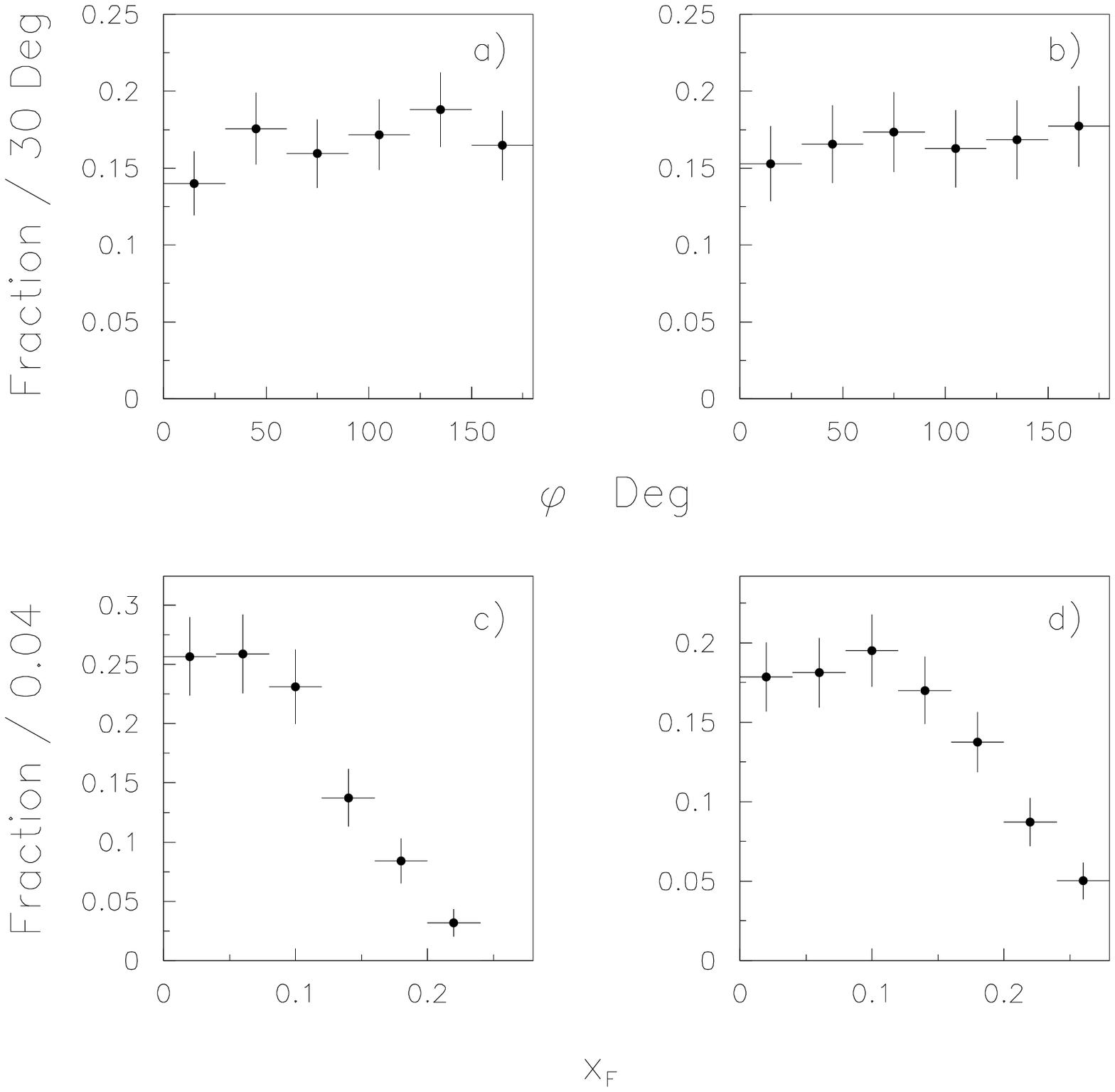,height=22cm,width=17cm}
\end{center}
\begin{center} {Figure 1} \end{center}
\newpage
\begin{center}
\epsfig{figure=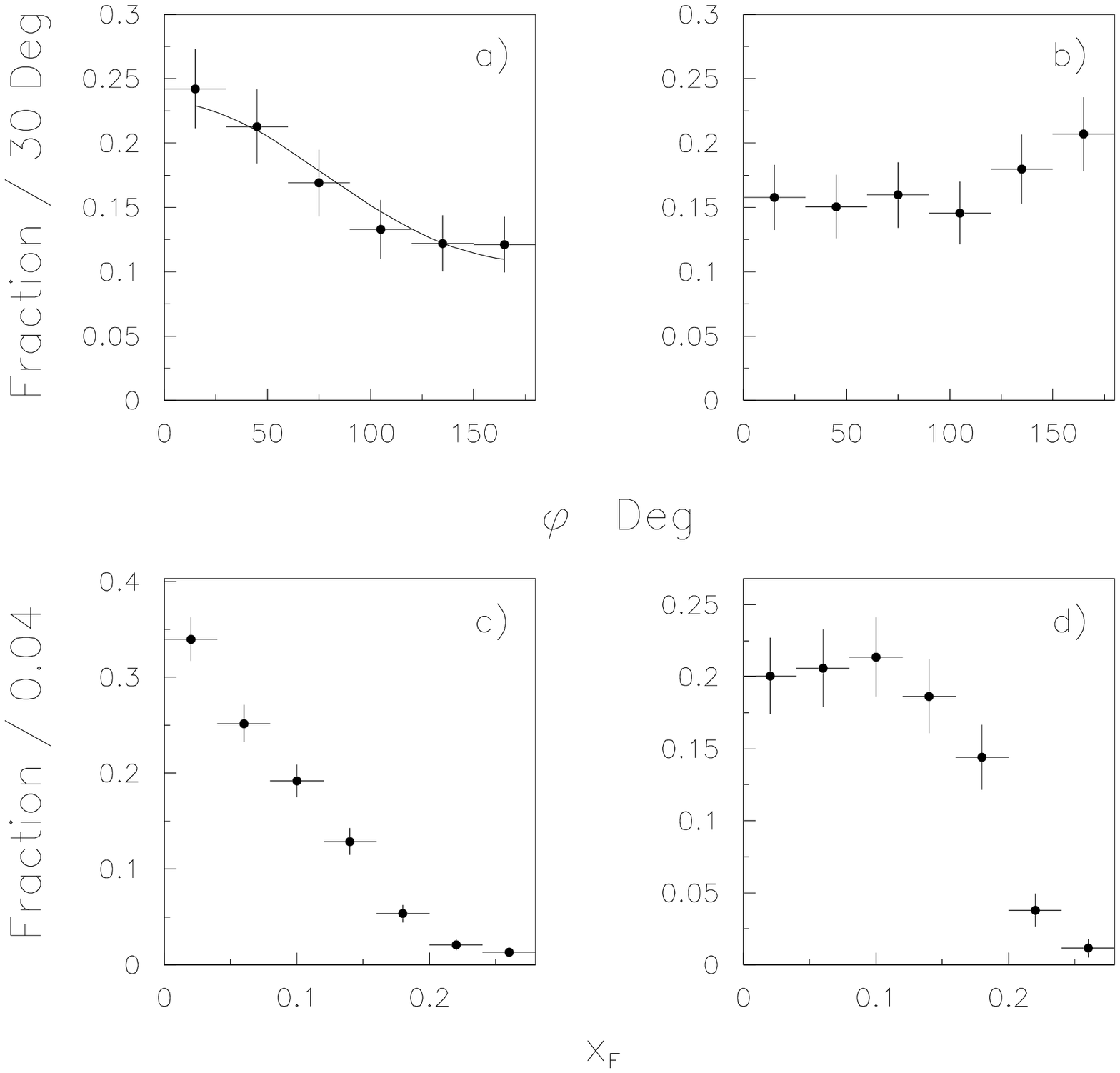,height=22cm,width=17cm}
\end{center}
\begin{center} {Figure 2} \end{center}
\newpage
\begin{center}
\epsfig{figure=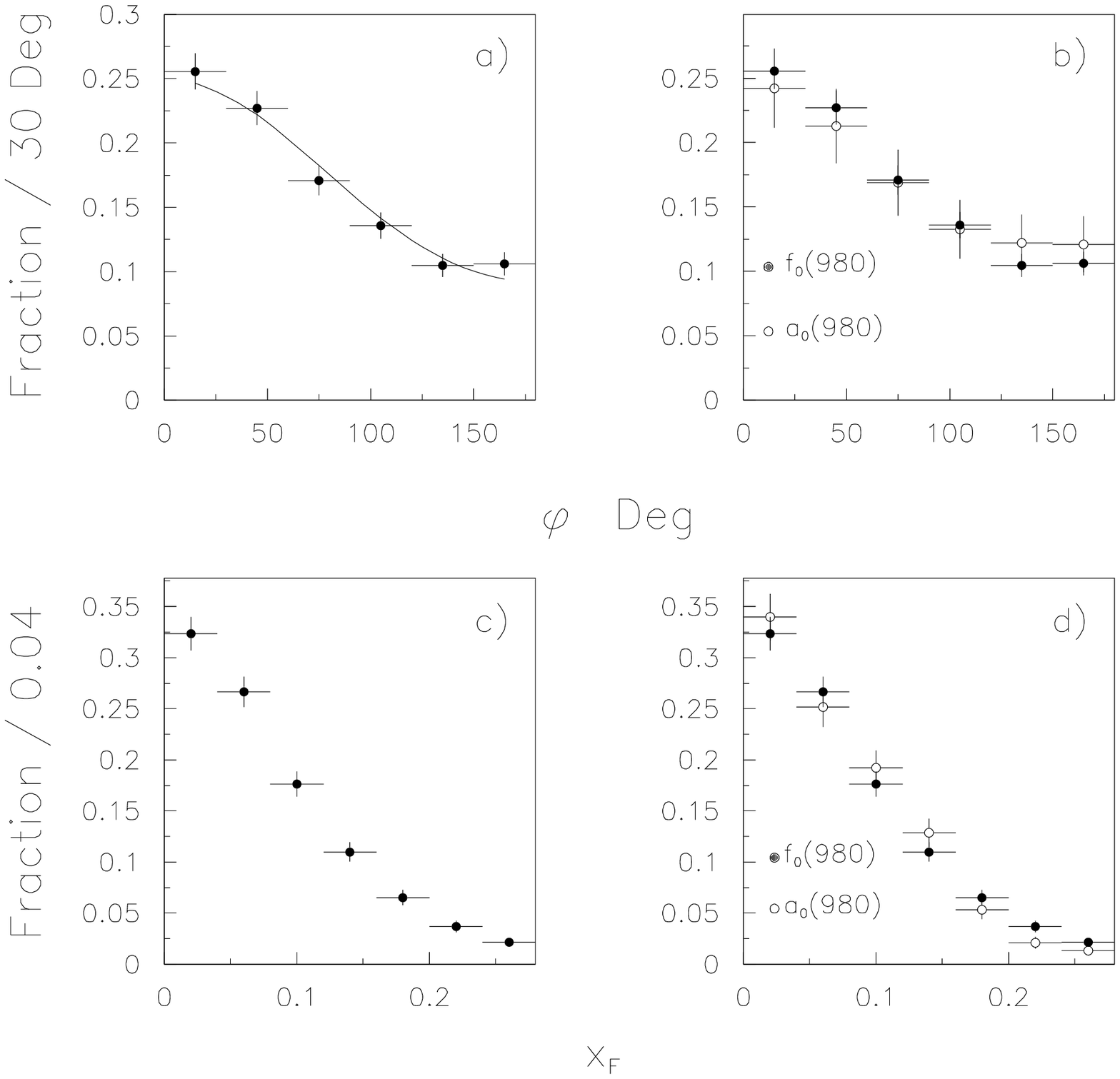,height=22cm,width=17cm}
\end{center}
\begin{center} {Figure 3} \end{center}
\end{document}